\documentclass[aps,prl,twocolumn,superscriptaddress,nofootinbib]{revtex4-1} 

\usepackage{epsfig}
\usepackage{color}
\usepackage[latin1]{inputenc}
\usepackage{float,amsmath}
\usepackage{graphicx}

\newcommand{\xt}{\mathbf{x}_T}
\newcommand{\be}{\begin{eqnarray}}
\newcommand{\ee}{\end{eqnarray}}

\newcommand{\benum}{\begin{enumerate}}
\newcommand{\eenum}{\end{enumerate}}

\usepackage{ulem}

\begin{document}

\title{Initial state geometry and the role of hydrodynamics in proton-proton, \\proton-nucleus 
and deuteron-nucleus collisions }

\author{Adam Bzdak}
\affiliation{RIKEN BNL Research Center, Brookhaven National Laboratory, Upton, NY 11973, USA}
\author{Bj\"orn Schenke}
\affiliation{Physics Department, Brookhaven National Laboratory, Upton, NY 11973, USA}
\author{Prithwish Tribedy}
\affiliation{Variable Energy Cyclotron Centre, 1/AF Bidhan Nagar, Kolkata 700064, India}
\author{Raju Venugopalan}
\affiliation{Physics Department, Brookhaven National Laboratory, Upton, NY 11973, USA}

\begin{abstract}
We apply the successful Monte Carlo Glauber and IP-Glasma initial state models of heavy ion collisions to the much smaller size systems produced in proton-proton, proton-nucleus and deuteron-nucleus collisions. We observe a significantly greater sensitivity of the initial state geometry to details of multi-particle production in these models compared to nucleus-nucleus collisions. In particular, we find that the size of the system produced in p+A collisions is very similar to the one produced in p+p collisions, and predict comparable Hanbury-Brown-Twiss radii in the absence of flow in both systems.
Differences in the eccentricities computed in the models are large, while differences amongst the generated flow coefficients $v_2$ and $v_3$ are smaller. For a large number of participants in proton-lead collisions, the $v_2$ generated in the IP-Glasma model is comparable to the value obtained in proton-proton collisions.
Viscous corrections to flow are large over characteristic lifetimes in the smaller size systems. In contrast, viscous contributions are significantly diminished over the longer space-time evolution of a heavy ion collision. 

\end{abstract}

\maketitle


\section{Introduction}
A recent discovery in high multiplicity proton-proton and proton-nucleus collisions are correlations between pairs of charged hadrons that are collimated in their relative azimuthal angle and are long range in relative rapidity. These ``ridge" correlations were mostly unanticipated for such small size systems, albeit a similar striking effect was previously seen in heavy ion collisions at RHIC and subsequently also at the LHC. The ridge in  $\sqrt{s}=7$ TeV proton-proton collisions was discovered by the CMS collaboration~\cite{Khachatryan:2010gv}. In proton-lead collisions at $\sqrt{s}=5.02$ TeV/nucleon, a sizable ridge was observed by the CMS collaboration~\cite{CMS:2012qk}, the ALICE  collaboration~\cite{Abelev:2012ola} and the ATLAS collaboration~\cite{Aad:2012gla}. In addition, the PHENIX collaboration at RHIC recently discovered a ridge in very central deuteron-gold collisions at $\sqrt{s}=200$  GeV/nucleon~\cite{Adare:2013piz}. 

These long range correlations are of fundamental importance because they probe the very early time dynamics of matter produced in hadronic collisions. A question of considerable recent interest is whether the ridge effect in p+p and p/d+A collisions is 
due to initial state effects arising from the correlations of gluons already present in the nucleon and nuclear wavefunctions, or whether it is due to final state rescattering effects that are amenable to a hydrodynamic description. In both cases, one assumes a dynamical scenario where long range rapidity correlations are generated in the initial state.~\footnote{In hydrodynamical models, these long range rapidity correlations are a consequence of the choice of initial conditions, wherein the initial transverse spatial profile of the energy density distribution is assumed to be the same at all rapidities. Though 
not widely appreciated, this choice corresponds to an assumption of strong long range correlations in the dynamics of multiparticle production at short transverse spatial distances. Only azimuthal correlations are dynamically generated by the hydrodynamic equations.}  
The question is whether the azimuthal collimation observed in the ridge is also due to the same initial state correlations that generate long range rapidity correlations, or whether they are generated primarily by the final state flow of these correlated structures. 

A powerful framework in which long range rapidity correlations can be computed systematically is the Color Glass Condensate (CGC) effective field theory (EFT) \cite{Gelis:2010nm}. In the CGC EFT, these correlations are a consequence of gluon saturation at central impact parameters in the proton and nuclear wavefunctions. In the hadronic collision, gluon fields are generated that stretch out in rapidity between the receding hadrons, and are coherent in the transverse plane over distances $1/Q_s$, where $Q_s$ is the saturation scale. The saturation scale in a hadron or nucleus is a function of the parton momentum fraction $x$ and impact parameter, and grows with increasing energy and nuclear size.  

Multi-particle production, by the decay of the gauge field configurations corresponding to these Glasma flux tubes~\cite{Dumitru:2008wn}, is nearly boost invariant and nearly azimuthally isotropic; the resulting multiplicity distribution is the negative binomial distribution~\cite{Gelis:2009wh}. The QCD graphs that generate these distributions are called ``Glasma graphs''. At high $k_T \gg Q_s$, the contribution of these graphs is highly suppressed. In contrast, for $k_T \leq Q_s$, where high occupancies in hadron wavefunctions are probed, Glasma graphs are enhanced by $\alpha_s^{-8}$, a factor of $\sim 10^5$ for typical values of the probed QCD fine structure constant $\alpha_s$. In nuclear collisions at ultra-relativistic energies, these (nearly) boost invariant configurations are argued to provide the dominant mechanism for multi-particle production, and factorization theorems (to leading logarithmic accuracy in $x$)
have been derived~\cite{Gelis:2008rw,Gelis:2008ad,Gelis:2008sz}. 

Though the bulk of multi-particle production is nearly azimuthally isotropic, it is not exactly so. As first noted in \cite{Dumitru:2010iy}, based on the formalism in  \cite{Gelis:2008sz,Dusling:2009ni}, Glasma graphs produce contributions that are collimated at relative azimuthal separations of $\Delta \Phi \approx 0$ and $\Delta \Phi \approx \pi$. It has been shown recently that these initial state contributions provide a quantitative description of the measured collimated yield in both proton-proton and proton-nucleus collisions~\cite{Dusling:2012iga,Dusling:2012cg,Dusling:2012wy,Dusling:2013oia}.

However, as observed previously~\cite{Voloshin:2003ud,Pruneau:2007ua}, long range rapidity correlations from the initial state can also be collimated by the radial flow of a fluid. Indeed, within the Glasma flux tube framework itself, the radial flow of Glasma flux tubes correlated over distance scales $1/Q_s$ was shown to generate a sizable ridge for large radial flow velocities~\cite{Dumitru:2008wn,Gavin:2008ev}. In nucleus-nucleus collisions, where large radial flow is generated, several groups have shown that hydrodynamical flow provides a very good explanation of the data on two-particle correlations in the $\Delta \eta$-$\Delta \Phi$ plane \cite{Takahashi:2009na,Luzum:2010sp,Gavin:2012zr,Bozek:2012en}. There have also been  attempts to extend this description of the ridge in nucleus-nucleus collisions to the ridges observed in high multiplicity proton-proton~\cite{Werner:2010ss,Avsar:2010rf} and p+Pb~\cite{Bozek:2011if,Bozek:2012gr,Bozek:2013df,Bozek:2013uha} collisions. In the latter case, it is claimed that features of LHC high multiplicity data on proton-nucleus collisions~\cite{Aad:2013fja} and corresponding data in deuteron-gold collisions at RHIC~\cite{Adare:2013piz} are quantitatively explained in the Monte-Carlo (MC)  Glauber hydrodynamic model of ~\cite{Bozek:2011if,Bozek:2012gr,Bozek:2013df}. 

We will argue here that the applicability of hydrodynamics to the smaller size systems of proton-proton and proton/deuteron-nucleus collisions is strongly dependent on assumptions about the nature of the initial multi-particle dynamics, much more so than 
in collisions of heavy nuclei. We will illustrate this by comparing results obtained in MC-Glauber models with particular dynamical assumptions about the initial state geometry with those obtained in the framework of the IP-Glasma initial state model~\cite{Schenke:2012wb,Schenke:2012hg} of hadrons and nuclei. Very noticeable differences are seen between the two models (with the same initial state configurations) for the computed eccentricities and corresponding flow coefficients. In contrast, both initial state models, when combined with event-by-event hydrodynamical simulations, as in \cite{Bozek:2012qs,Schenke:2010nt,Schenke:2010rr,Schenke:2011bn,Gale:2012rq}, give similarly good descriptions\footnote{The IP-Glasma+\textsc{music} model of ~\cite{Gale:2012rq} also reproduces the event-by-event $v_n$ fluctuations measured by the ATLAS collaboration~\cite{Jia:2012ve}; at present, it appears to be the only model that successfully reproduces these flow fluctuations.} of bulk multiplicity and flow observables in heavy-ion collisions at both RHIC and the LHC. 

The paper is organized as follows. In the next section, we will outline the different methods employed to compute the initial spatial sizes and eccentricities and some of the consequences thereof. We will review the IP-Glasma model, and show its predictions for the initial spatial sizes in proton-proton and proton-nucleus collisions. We will compare the eccentricities obtained in this model to those in various implementations of the MC-Glauber model for proton-nucleus. The generated flow in proton-proton and proton/deuteron-nucleus collisions is considered next and contrasted between the two models. The final section   discusses the magnitude of viscous effects in different implementations of viscous hydrodynamics in proton-nucleus and nucleus-nucleus collisions. We end with a brief summary and outlook. 

\section{Models of the initial state geometry}

Modeling  the initial state in p+A, d+A and especially p+p collisions is 
a lot more challenging than in A+A collisions. In the latter, the system's geometry is primarily characterized by the overall 
shape of the interaction region. The dominant component in shape fluctuations are due to geometrical fluctuations of nucleon positions inside 
the nuclei from event-to-event. The large number of participants allows one, to first approximation, to neglect the dynamical details 
of how energy is deposited in A+A interactions. In p+A and d+A collisions,  the system's geometry is very sensitive to the 
proton (or deuteron) size, and the detailed nature of multi-particle production and the spatial distribution 
of the produced energy density become important. In particular, sub-nucleon size fluctuations (with
characteristic length scales less than $1\,{\rm fm}$) contribute significantly to the initial geometry of matter produced in the collision. 

\begin{figure}[ht]
  \centering
  \includegraphics[width=7cm]{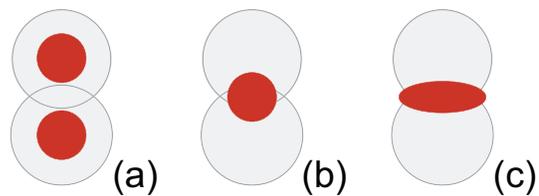}
  \caption{(Color online) Various models of the energy density deposition (denoted by red dots) in nucleon-nucleon
  collisions. In the left plot (a) the energy density is produced at the center of the colliding nucleons even for grazing
  collisions. 
  The center and right plots (b),(c) correspond to different eccentricities depending on the matter distribution in the nucleon
  overlap region. 
  For the configuration depicted on the left eccentricity $\varepsilon_2 \approx 1$, whereas for the configuration in the center $\varepsilon_2=0$.} 
  \label{fig:3models} 
\end{figure}

The spatial eccentricities that characterize the geometry of the initial state can be defined as 
\begin{equation}\label{eq:en}
  \varepsilon_n = \frac{\sqrt{\langle r^n \cos(n\phi)\rangle^2+\langle r^n \sin(n\phi)\rangle^2}}{\langle r^n \rangle}\,,
\end{equation}
where $\langle \cdot \rangle$ is either an average over all participant nucleon positions characterized by the nucleon centers or an average weighted by the deposited energy density.

In each realization of the MC-Glauber model, all participants contribute equally to the energy density 
deposited in the system. For example, two nucleons that barely touched each other in the collision 
are assumed to deposit the same amount of energy as two nucleons that interacted with zero relative impact parameter. However, in a microscopic parton model based picture of the hadron, it is natural to expect that peripheral nucleon-nucleon collisions deposit significantly less matter than the central ones. This is because the parton density decreases rapidly with impact parameter~\cite{Mueller:1993rr}, a picture confirmed by phenomenological descriptions of diffractive deeply inelastic scattering data from the HERA collider in the framework of the IP-Sat dipole model~\cite{Kowalski:2003hm}. This physics is naturally incorporated in the IP-Glasma approach, with the saturation scale depending on the impact parameter \cite{Schenke:2012wb}. 

In many Glauber model computations, the energy even in a peripheral collision is deposited in the center of the wounded nucleons \cite{Bialas:1976ed}. This is sketched in the left most configuration shown in Fig.~\ref{fig:3models}, and corresponds to the ellipticity $\varepsilon_2 \approx 1$ and triangularity $\varepsilon_3 =0$, when participant centers are used to compute the average in Eq.\,(\ref{eq:en}). However, on the basis of the parton model arguments outlined, one expects that peripheral nucleon-nucleon collisions will deposit most of the produced energy in the region of overlap, as illustrated in the middle figure of Fig.~\ref{fig:3models}. In this case, both the ellipticity $\varepsilon_2=0$ and the triangularity $\varepsilon_3=0$, if one assumes an isotropic energy density deposition. Following the shape of the overlap region for the deposition of energy density, as sketched in the right most figure in Fig.~\ref{fig:3models}, can produce instead eccentricities that are anywhere in between the two extremes of the other two configurations.

In A+A collisions, because of the large number of overlapping nucleons, these finer details of geometry are less important. In contrast, these different microscopic pictures will have significant consequences for multiparticle production in nucleon-nucleon and nucleon-nucleus collisions. We note that in the IP-Sat framework, where particle production only occurs in the geometrical overlap regions, a good description is obtained of the n-particle multiplicity distributions in proton-proton collisions at central rapidities from 200 GeV to 7 TeV~\cite{Tribedy:2010ab,Tribedy:2011aa}. 

We will now outline the IP-Glasma model, which has been discussed at length elsewhere \cite{Schenke:2012wb,Schenke:2012hg}. In this 
model, which goes significantly beyond the treatment of hadron collisions in \cite{Tribedy:2010ab,Tribedy:2011aa}, incoming gluon fields are computed from fluctuating color charges via the classical QCD Yang-Mills equations. In an individual nucleon, the color charges are assumed to follow a local Gaussian distribution with variance $g^2\mu_p^2(x,\xt)$, a quantity proportional to the saturation scale $Q_s^{(p)}(x,\xt)$ in the proton. This latter quantity is determined from the IP-Sat dipole model \cite{Kowalski:2003hm} with parameters fit to HERA data on inclusive and exclusive final states \cite{Rezaeian:2012ji}.  Products of the nucleon dipole S-matrices, generate lumpy configurations of glue in nuclei, and are in agreement with extant fixed target data on electron-nucleus scattering~\cite{Kowalski:2007rw}. This gives the variance $g^2 \mu_A^2$ of Gaussian distributed  nuclear charge distributions. The latter, in the MV model~\cite{McLerran:1993ni,McLerran:1993ka}, is used to solve for the coherent classical gauge fields of the nuclei before the collision. Here, as opposed to the previous implementation described in \cite{Schenke:2012wb,Schenke:2012hg}, we compute $x = Q_s(x,\xt)/\sqrt{s}$ self-consistently at every transverse position $\xt$. Further, the running coupling is evaluated locally as $\alpha_s({\bar Q}_S(x,\xt))$, where 
${\bar Q}_S = {\rm max} (Q^A_S, Q^B_S)$ where $Q^{A(B)}_S$ are the saturation scales of the projectile (target).

The solution for the transverse (longitudinal) gauge fields after the collision is obtained in Schwinger gauge $A^\tau=0$ and is given by the sum (commutator) of the incoming purely transverse fields to determine the initial energy and number content of the Glasma fields~\cite{Kovner:1995ja,Kovchegov:1997ke,Krasnitz:1998ns,Krasnitz:1999wc,Krasnitz:2000gz,Lappi:2003bi}. The gauge fields are regulated by an infrared mass scale $m=0.1$ GeV, which ensures the unphysical Coulomb tail from solutions of the Yang-Mills equations is suppressed at large distances. This procedure, and the fact that the incoming fields are pure gauges, ensures that no energy density will be deposited in regions where at least one of 
the incoming fields vanishes. This key property of the solutions determines that in p+A (d+A) collisions the system size is 
dominated by the size of the incoming proton (and neutron). 

\begin{figure}[htb]
   \begin{center}
     \includegraphics[width=8.5cm]{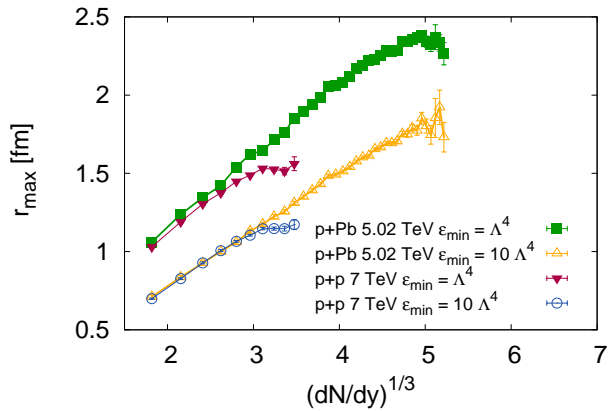}
     \caption{(Color online) System size in p+p and p+Pb collisions as 
     a function of the number of gluons to the power of $1/3$, $(dN/dy)^{1/3}$, computed in the IP-Glasma model. 
     $r_{\rm max}$ is the maximal radius for which the energy density of the Yang-Mills fields are above a minimal 
     value of $\varepsilon_{\rm min}=\alpha \Lambda_{\rm QCD}^4$,  
     with $\alpha=1$ (filled symbols) and $\alpha=10$ (open symbols). 
     Note different energies for p+p and p+Pb. }
     \label{fig:r}
   \end{center}
\end{figure}

In Fig.\,\ref{fig:r}, the system size $r_{\rm max}$, the maximal radius for which the energy density of the Yang-Mills fields is above a minimal value of $\varepsilon=\alpha \Lambda_{\rm QCD}^4$ (with $\alpha \in \{1,10\}$), is presented. 
We observe that the system sizes in p+p and p+Pb are comparable and grow approximately linearly as a function of the 
number of gluons to the power of $1/3$. Rather than the difference between p+p and p+Pb, the result is more sensitive to the choice of the value of $\alpha$, albeit relatively weakly given that the energy density changes by an order of magnitude. This uncertainty represents intrinsic non-perturbative effects that cannot be further quantified within our present knowledge of QCD. 

Our observation of the scaling of the system sizes in p+p and p+Pb collisions indicates that their HBT 
radii~\footnote{It is interesting to note that a very similar trend to that of our $r_{\rm max}$ is observed 
in p+p data on HBT radii measured at LHC \cite{Aamodt:2011kd}, where all HBT radii grow approximately linearly with the number of produced particles to the power of $1/3$. } should be comparable in value.
Indeed, for the same number of produced particles and comparable sizes of two systems, leading to comparable energy densities, both systems are very similar and we do not expect to observe significant differences in their HBT radii. Let us note that we do not calculate explicitly HBT radii in p+p and p+Pb, we only compare both systems based on their sizes and energy densities. If subsequently fluid dynamical evolution in p+A collisions is significant relative to p+p collisions, we would then anticipate significantly different HBT radii in p+Pb, as recently discussed in Ref. \cite{Bozek:2013df}. Thus HBT radii will help to discriminate between models of the spatial distribution of matter in the initial state and the magnitude of radial flow experienced in each.

\begin{figure}[htb]
  \begin{center} 
     \includegraphics[width=8cm]{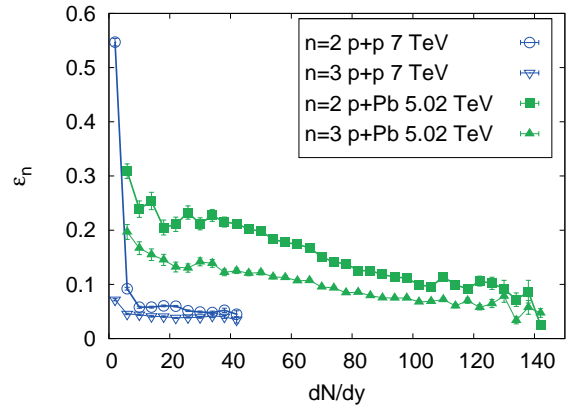}
     \caption{(Color online) Eccentricities $\varepsilon_2$ and  $\varepsilon_3$ 
     as a function of $dN/dy$ in the IP-Glasma model for p+Pb (filled symbols) and p+p (open symbols) collisions. }
     \label{fig:ecc}
   \end{center}
\end{figure}

In Fig. \ref{fig:ecc}, the ellipticity and triangularity computed in the IP-Glasma model are plotted as a function of  $dN/dy$, the gluon number per unit rapidity in the model. The ellipticities in p+p collisions are significantly lower than in p+Pb collisions, except at very low values of the multiplicity.
The triangularity $\varepsilon_3$ in p+p collisions is very small and is comparable to the ellipticity. 
The triangularity in p+Pb collisions is consistently larger than in p+p, though distinctly smaller than 
the ellipticity in p+Pb up to very high multiplicities. 

\begin{figure}[htb]
   \begin{center}
     \includegraphics[width=9cm]{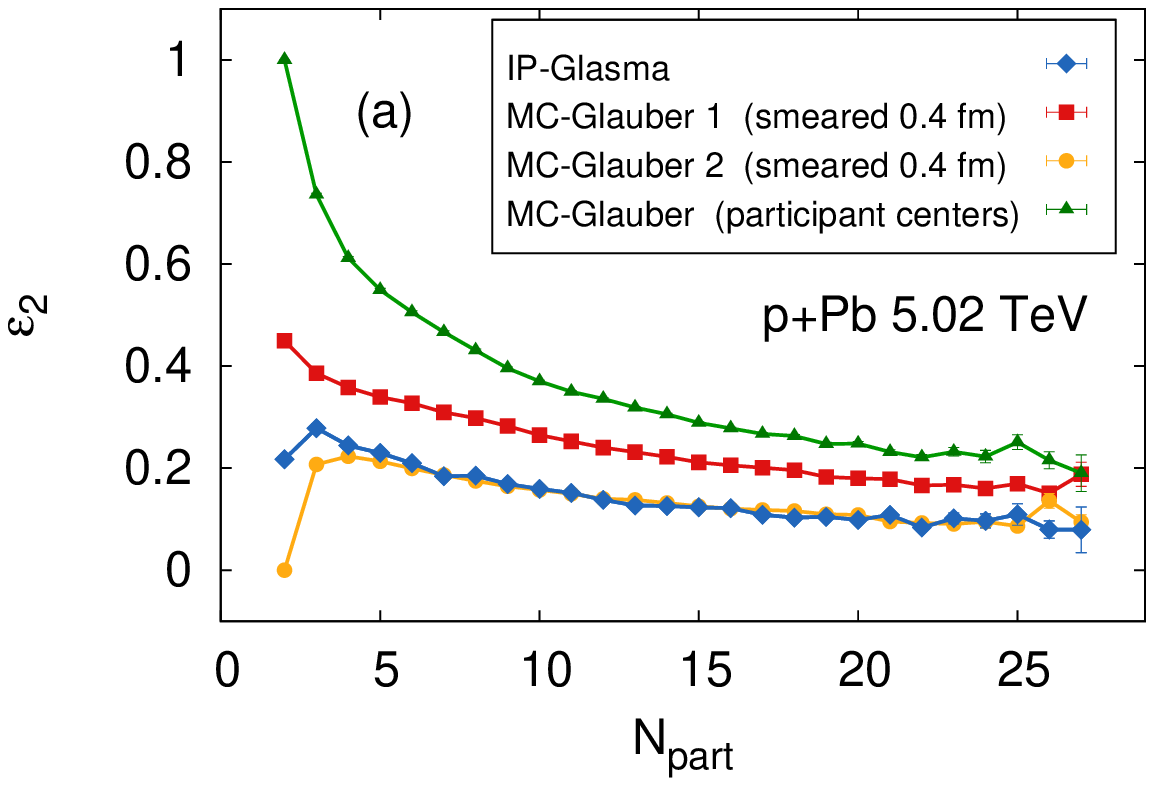}\\
     \includegraphics[width=9cm]{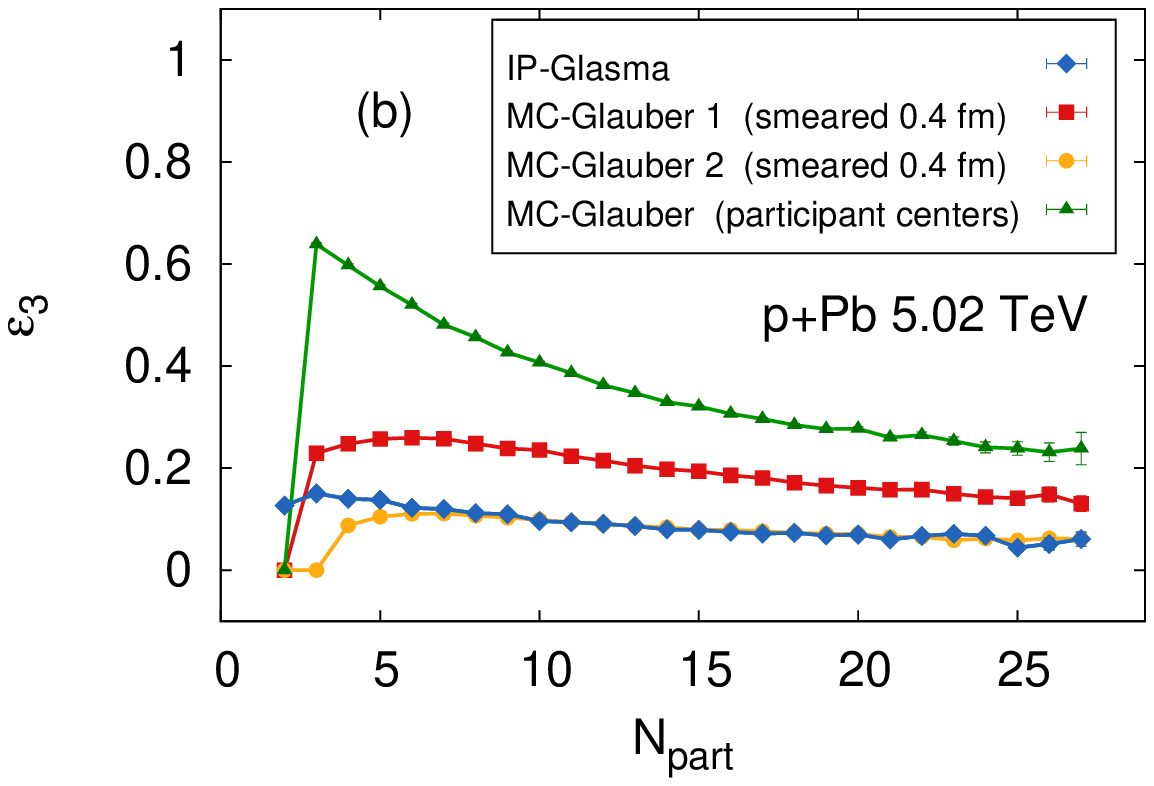}
     \caption{(Color online) Eccentricities $\varepsilon_2$ (a) and  
     $\varepsilon_3$ (b) as a function of the number of wounded nucleons $N_{\rm part}$. 
     In the MC-Glauber (participant centers) model the energy density is deposited in the centers of wounded nucleons (without smearing).
     Smearing energy densities with the Gaussian distribution ($\sigma_0=0.4\,{\rm fm}$) results in the 
     MC-Glauber 1 model. In the MC-Glauber 2 model the energy density is smeared about the midpoint between colliding 
     nucleons.}  
     \label{fig:eccNpart}
   \end{center}
\end{figure}

In Fig. \ref{fig:eccNpart}, we plot $\varepsilon_2$, $\varepsilon_3$ in proton-lead collisions in the IP-Glasma model as a function of the number of participants $N_{\rm part}$ and compare the results to two different realizations of the Monte-Carlo Glauber model.
Computing eccentricities using participant centers in a MC-Glauber model, as done in \cite{Bozek:2011if}, the results are along the lines anticipated in our discussion of Fig.~\ref{fig:3models}. $\varepsilon_2$ is exactly unity and $\varepsilon_3=0$ for $N_{\rm part} =2$. 
A similar trend for $\varepsilon_2$ at low $N_{\rm part}$ is observed when computing eccentricities using Gaussian energy densities (with width $\sigma_0=0.4\,{\rm fm}$) in the centers of participants (MC-Glauber 1) as a weight. In this case the eccentricities are noticeably lower. The MC-Glauber 1 realization of the Glauber model is similar to the one employed in the computations of ~\cite{Bozek:2011if,Bozek:2012gr,Bozek:2013df} for proton-nucleus and deuteron-nucleus collisions. 

Assigning a Gaussian distributed energy density to the midpoint between two colliding nucleons (MC-Glauber 2), as illustrated in the middle figure in Fig.\,\ref{fig:3models}, reduces the resulting eccentricities significantly, with the difference in $\varepsilon_2$ between models 1 and 2 being approximately a factor of 2. We have checked that decreasing the smearing width from $\sigma_0=0.4\,{\rm fm}$ increases both eccentricities.

We have used a black disk approximation of the cross-section, meaning that nucleons are wounded whenever their geometric distance from a nucleon of the other nucleus is less than $r_{NN} = \sqrt{\sigma_{NN}/\pi}$, where $\sigma_{NN}$ is the nucleon-nucleon inelastic cross section. Alternatively, one can introduce a smooth profile of the nucleon that determines the probability for an interaction at a given nucleon-nucleon distance. This profile can be extracted from the p+p differential elastic cross section data \cite{Amaldi:1979kd} and can be approximated by a Gaussian \cite{Bialas:2006kw,Bialas:2006qf,Rybczynski:2011wv}. Its use has been argued to be preferable because it does not result in extremely large elastic nucleon-nucleon cross sections as the hard-sphere case. Using a smooth profile, we find an increase in both the system size (by up to 50\%) and the eccentricities (by up to a factor of 2). Again, it demonstrates that eccentricities in p+Pb collisions are very sensitive to details of nucleon-nucleon interactions. 

In the IP-Glasma model, both the ellipticity and triangularity coincide with model MC-Glauber 2 for all but the smallest values of $N_{\rm part}$. This agreement is however a coincidence for the value of $\sigma_0=0.4$ fm chosen in the MC-Glauber model and will not hold if this parameter is varied. In the IP-Glasma model, there is no such free parameter. 

The differences between the MC-Glauber realizations used in hydrodynamical models and the IP-Glasma model are strikingly seen in deuteron-gold collisions. For the deuteron, we use the Hulthen form of the wave function, 
\begin{equation}
  \phi_{\rm pn}(r) = \frac{1}{\sqrt{2\pi}}\frac{\sqrt{a b(a+b)}}{b-a} 
  \frac{e^{-a r} - e^{-b r}}{r}\,,
\end{equation}
with $r$ being the distance between the proton and the neutron and the parameters $a=0.228\,{\rm fm^{-1}}$ 
and $b=1.18\,{\rm fm^{-1}}$ \cite{Miller:2007ri}.
Integrating $\phi^2_{\rm pn}$ over $z=\sqrt{r^2-r_T^2}$, where $r_T$ is the distance between 
the proton and the neutron in the projection onto the
transverse plane, we obtain the thickness function $T_d(r_T)$.

\begin{figure}[ht]
  \centering
  \includegraphics[width=6.5cm]{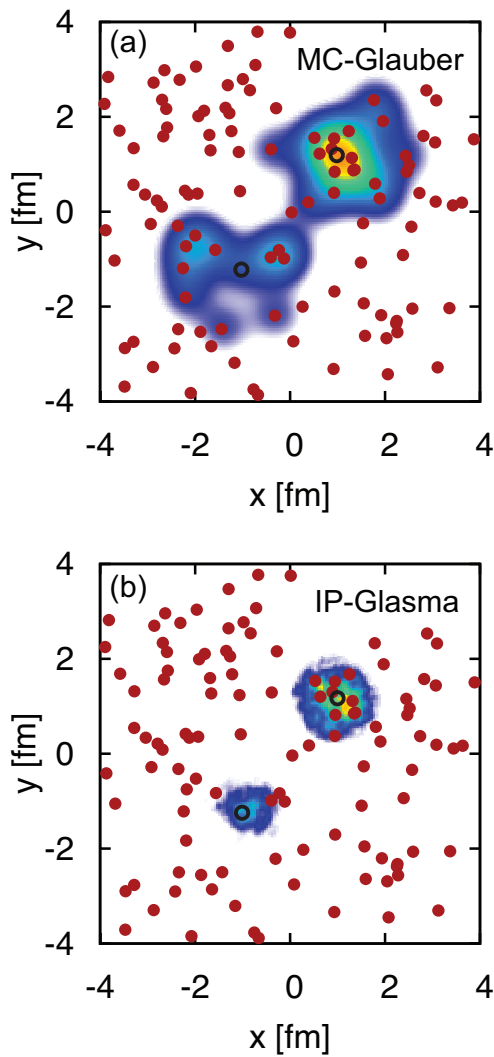}
  \caption{(Color online) Initial energy density distribution (arbitrary units, increasing from blue to red) in 
  the transverse plane in a d+Au collision in the MC-Glauber model (a)
  and the IP-Glasma approach (b). The nucleon positions (open circles for the deuteron, solid circles 
  for gold) are exactly the same in the two cases.}
  \label{fig:dAu-eps}
\end{figure}

In Fig.~\ref{fig:dAu-eps}, we show plots with a typical deuteron configuration in the transverse plane (denoted by open circles) superposed on transverse projections of the nucleon positions in the gold nucleus (denoted by filled circles). In the top plot, we show the energy density contours 
from the MC-Glauber 1 model and in the bottom plot, the corresponding IP-Glasma model results. These are seen to be quite different. In the latter, it is observed that the peaks in the contour are closely associated with the centers of the deuteron nucleon positions and vary strongly depending on the number of gold nucleon positions in their immediate vicinity. In the former MC-Glauber case, significant energy densities are seen even in regions where nucleons of the gold nucleus are widely separated in transverse spatial position from the deuteron nucleons. Nucleons that have been marginally grazed produce as much energy density as those that have suffered a head on collision. In the IP-Glasma model, because the mean distance in the projection onto the transverse plane between the two nucleons in a deuteron is $2.52\,{\rm fm}$, the majority of events have widely separated interaction regions. This is quite different in the MC-Glauber model. 

Whether eccentricity is a relevant measure in deuteron-gold collisions depends sensitively on the radial separation of the regions where energy density is deposited. If they are too far apart for hydrodynamic flow to bring them into contact over the system's lifetime, the eccentricity will be a poor measure of flow. If they are close enough at the same eccentricity to influence subsequent flow, the eccentricity will track flow better. Thus eccentricity in deuteron-gold collisions, in contrast to nucleus-nucleus collisions, is at best a qualitative measure of anisotropic flow.
 
\section{Flow in proton-proton and proton/deuteron-nucleus collisions}

An interesting compilation of the ratio of the elliptic flow coefficient to the ellipticity $v_2/\varepsilon_2$ 
versus the multiplicity at central rapidity in proton-nucleus, deuteron-nucleus, and a variety of centralities in nucleus-nucleus collisions can be found in \cite{Adare:2013piz}. This scaling may be taken to suggest the same pattern of collective flow in the smaller size systems as in the larger size one. However, as noted, the effect of the different eccentricity computations on $v_2/\varepsilon_2$ for p/d-A collisions is dramatic because $\varepsilon_2$ in the IP-Glasma model is two to five times smaller than the MC-Glauber model depending on $N_{\rm part}$. For A+A collisions, the differences in the eccentricity computations are not as large \cite{Schenke:2012wb}, and the scaling of $v_2/\varepsilon_2$ is less sensitive to model assumptions.

In addition, from the point of view of examining the presence of collective flow in the system, a more useful variable to plot $v_2/\varepsilon_2$ against is  the multiplicity per unit transverse overlap area $S_\perp$ \cite{Bhalerao:2005mm}. Such a compilation, for a wide range of centralities in heavy-ion collisions at the LHC and RHIC can be found in \cite{Chatrchyan:2012ta}. With this criterion however, when the x-axis is divided by $S_\perp$, the scaling of $v_2/\varepsilon_2$ is broken immediately for d+Au collisions. Indeed, the initial transverse area in d+Au is approximately a factor of two larger than in p+A collisions.  It will be interesting to calculate the transverse area for all colliding systems systematically in different models and investigate further the model dependence of scaling of $v_2/\varepsilon_2$ in smaller size systems. This study will be reported elsewhere. 

To study whether the behavior of the eccentricities presented in Fig.\,\ref{fig:eccNpart} is a good representation of the generated flow if the system
behaves hydrodynamically, we compute the root mean square $v_2$ and $v_3$ integrated over $p_T>0.5\,{\rm GeV}$ for different $N_{\rm part}$ in 
the MC-Glauber 1 and IP-Glasma model. We use a constant $\eta/s=0.08$, an initial time of hydrodynamic evolution $\tau_0=0.2\,{\rm fm}$ and a freeze-out temperature of $T_{\rm fo} = 120\,{\rm MeV}$. The normalization of the initial energy density in the IP-Glasma model was tuned to reproduce the charged particle multiplicity in p+p collisions at $7\,{\rm TeV}$. 
In the MC-Glauber 1 model the normalization of the energy density was set to approximately produce the same amount of charged particles as the IP-Glasma model. Because we are only interested in general trends in this work and not a detailed comparison to experimental data, we have not performed any fine tuning of parameters to reproduce particle spectra in p/d+A collisions. The reader should note however that these initial conditions, specifically the very low $\eta/s$ and small $\tau_0$, can reasonably be considered to provide upper bounds on the magnitude of the generated 
flow.\footnote{For nucleus-nucleus collisions at RHIC and the LHC, average values of $\eta/s=0.12$ and $\eta/s=0.2$ respectively, give the best fits to data~\cite{Gale:2013da}.}

\begin{figure}[htb]
   \begin{center}
     \includegraphics[width=8cm]{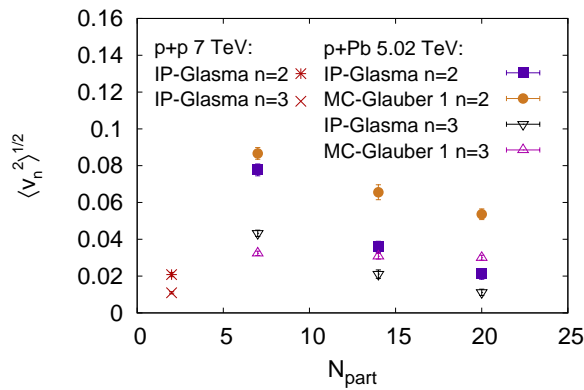}
     \caption{(Color online) Integrated $\langle v_2^2 \rangle^{1/2}$ and $\langle v_3^2 \rangle^{1/2}$ for charged hadrons in p+Pb collisions at different $N_{\rm part}$ in the MC-Glauber 1 and IP-Glasma model for $p_T>0.5\,{\rm GeV}$ and $\eta/s=0.08$. $v_2$ decreases with $N_{\rm part}$. Results for p+p collisions are for $b=0\,{\rm fm}$.  }
     \label{fig:v2tot-pPb}
   \end{center}
\end{figure}

\begin{figure}[htb]
   \begin{center}
     \includegraphics[width=8cm]{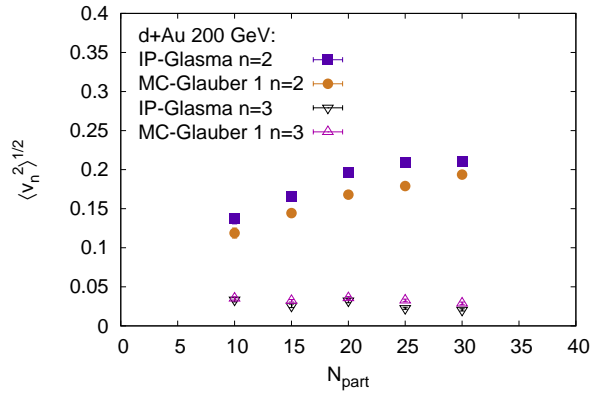}
     \caption{(Color online) Integrated $\langle v_2^2 \rangle^{1/2}$ and $\langle v_3^2 \rangle^{1/2}$ for charged hadrons in d+Au collisions at different $N_{\rm part}$ in the MC-Glauber 1 and IP-Glasma model for $p_T>0.5\,{\rm GeV}$ and $\eta/s=0.08$. $v_2$ increases with $N_{\rm part}$. }
     \label{fig:v2tot-dAu}
   \end{center}
\end{figure}

We show results for the integrated root mean square $v_2$ and $v_3$ 
for p+Pb collisions in Fig.\,\ref{fig:v2tot-pPb} and for d+Au collisions in Fig.\,\ref{fig:v2tot-dAu}. The first thing to note is the qualitative difference between the centrality dependence of $v_2$ in p+Pb and d+Au collisions. While in p+Pb collisions $v_2$ drops with increasing $N_{\rm part}$ as expected from $\varepsilon_2(N_{\rm part})$ in d+Au we find the opposite behavior. This behavior is expected qualitatively from $\varepsilon_2(N_{\rm part})$ in d+Au collisions \cite{Bozek:2011if}. However, as per our discussion in the previous section,  
$\varepsilon_2$ alone is not necessarily useful for a quantitative understanding of flow in d+Au collisions. 

In p+Pb collisions, given the eccentricity $\varepsilon_2$ of the MC-Glauber 1 model in Fig.\,\ref{fig:eccNpart},
one might naively expect an increase of $v_2$ by a factor of three when going from $N_{\rm part}=20$ to $N_{\rm part}=7$ if it scales with $\varepsilon_2$. While we do find the same trend, $v_2$ changes by a relatively smaller factor of approximately $1.7$. In the IP-Glasma model, the change in $v_2$  with $N_{\rm part}$ is larger (a factor of 2.5) even though the eccentricity $\varepsilon_2$ varies more slowly than in the MC-Glauber 1 model. For $N_{\rm part} =14$, $v_2$ is approximately a factor of two smaller than in the MC-Glauber 1 model, and about 60\% lower for $N_{\rm part} = 20$. 
$v_3$ is nearly flat in the MC-Glauber 1 model and decreases with $N_{\rm part}$ in the IP-Glasma model for both p+Pb and d+Au collisions.

We conclude that for small size systems, like p+Pb or d+Au, there is no simple quantitative scaling of the flow with eccentricity. Further, a smaller ellipticity in a different initial state model does not necessarily lead to smaller $v_2$ in that model, because the geometries (and system sizes) may be so drastically different, that $\varepsilon_2$ is not a sufficient predictor of $v_2$. This is seen strikingly for the flow generated in p+p collisions relative to p+Pb collisions. On the basis of the plots in Fig.~\ref{fig:ecc}, one might conclude that flow is much smaller in p+p relative to p+A. 

We computed the integrated ($p_T>0.5\,{\rm GeV}$) anisotropic flow for p+p collisions at $\sqrt{s}=7\,{\rm TeV}$ in the IP-Glasma model. 
We find $\langle v_2^2 \rangle^{1/2} \approx 0.02$ at $b=0\,{\rm fm}$ and  $\langle v_2^2 \rangle^{1/2}\approx 0.035$ at $b=1\,{\rm fm}$ --
for $b=0\,{\rm fm}$ the multiplicities in p+p are typically large, and the results can be qualitatively compared to those in p+Pb collisions. As shown in Fig.\,\ref{fig:v2tot-pPb}, $v_2$ at $N_{\rm part}=20$ in p+Pb is comparable to the results in p+p collisions within 50\%. The latter values, computed in the IP-Glasma model, are given by the points at $N_{\rm part}=2$ in Fig.\,\ref{fig:v2tot-pPb}.
We further find $\langle v_3^2 \rangle^{1/2}\approx 0.01$ for both studied impact parameters in p+p collisions, to be compared to the values shown for p+Pb in Fig.\,\ref{fig:v2tot-pPb}.

\begin{figure}[htb]
   \begin{center}
     \includegraphics[width=8cm]{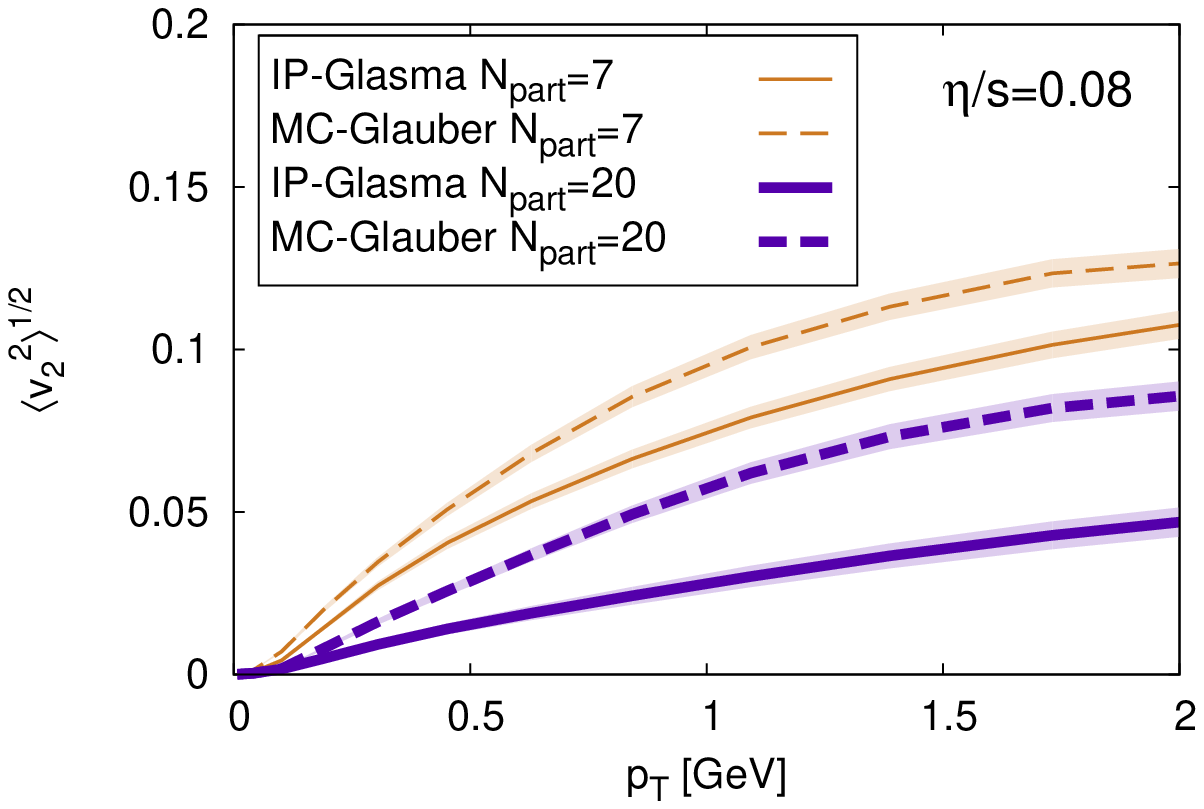}
     \caption{(Color online) $v_2(p_T)$ for charged hadrons in p+Pb collisions at fixed $N_{\rm part}=7$ (thin lines) and 20 (thick lines) in the MC-Glauber (dashed) and IP-Glasma (solid) model.}
     \label{fig:v2-fixedNpart}
   \end{center}
\end{figure}

\begin{figure}[htb]
   \begin{center}
     \includegraphics[width=8cm]{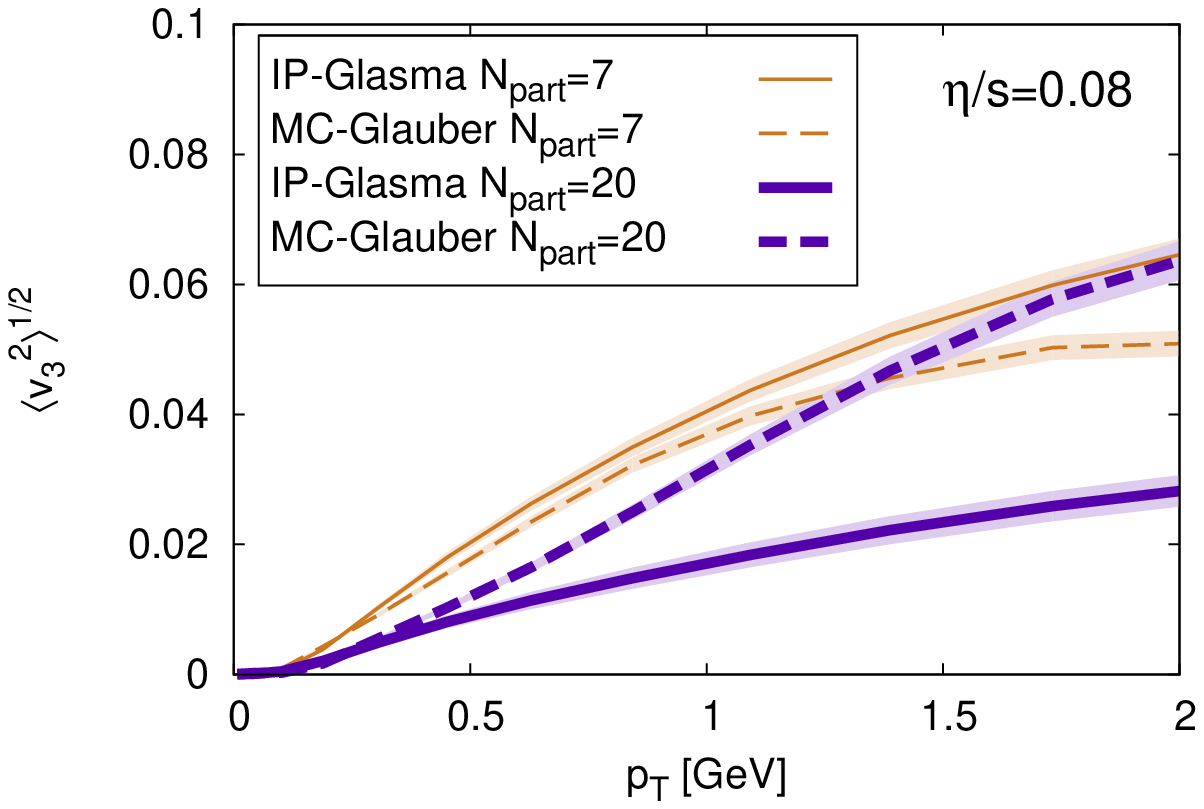}
     \caption{(Color online) $v_3(p_T)$ for charged hadrons in p+Pb collisions at fixed $N_{\rm part}=7$ (thin lines) and 20 (thick lines) in the MC-Glauber (dashed) and IP-Glasma (solid) model.}
     \label{fig:v3-fixedNpart}
   \end{center}
\end{figure}

\begin{figure}[htb]
   \begin{center}
     \includegraphics[width=8cm]{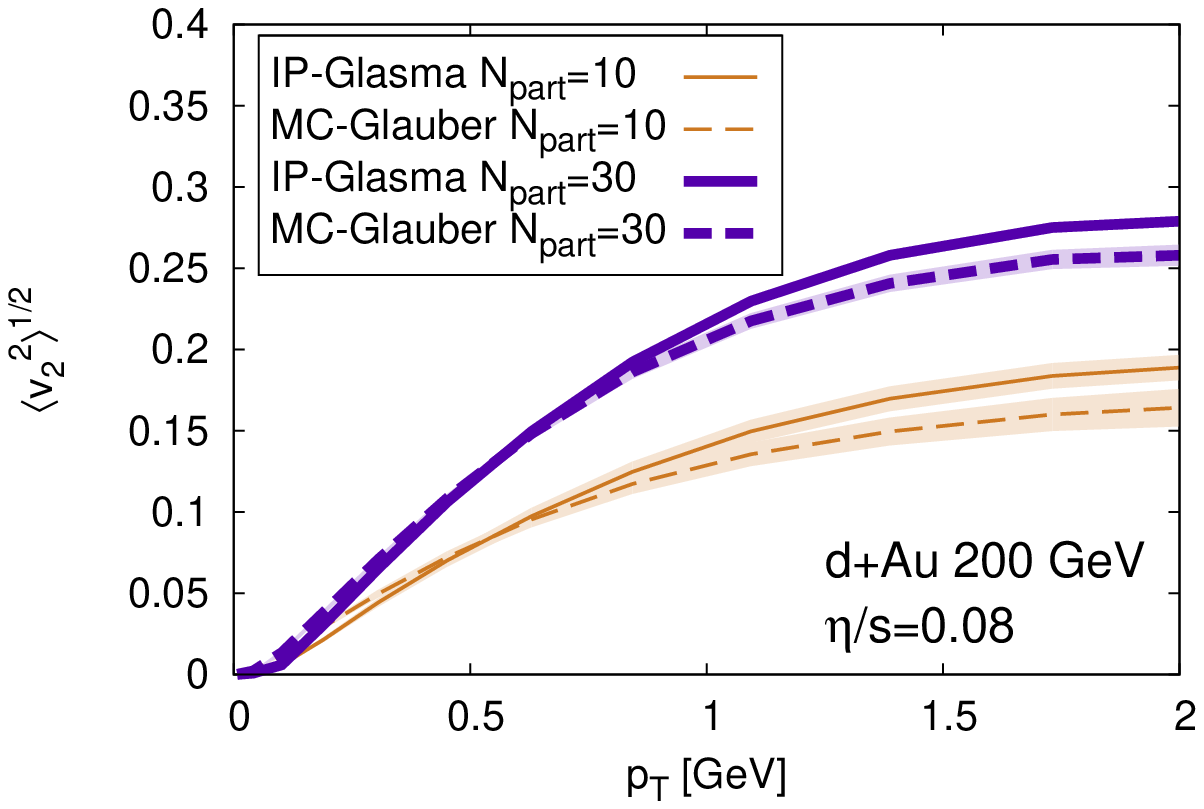}
     \caption{(Color online) $v_2(p_T)$ for charged hadrons in d+Au collisions at fixed $N_{\rm part}=10$ (thin lines) and 30 (thick lines) in the MC-Glauber (dashed) and IP-Glasma (solid) model.}
     \label{fig:v2-fixedNpart-dAu}
   \end{center}
\end{figure}

\begin{figure}[htb]
   \begin{center}
     \includegraphics[width=8cm]{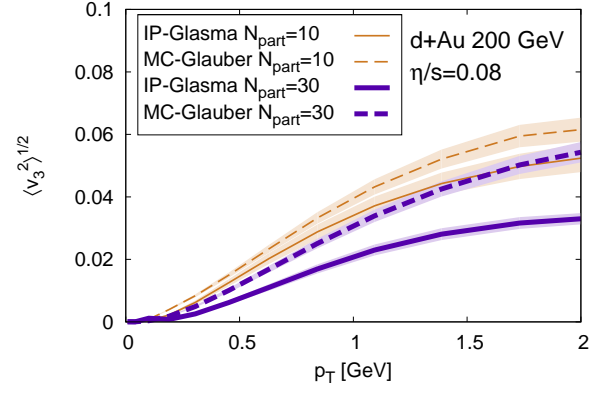}
     \caption{(Color online) $v_3(p_T)$ for charged hadrons in d+Au collisions at fixed $N_{\rm part}=10$ (thin lines) and 30 (thick lines) in the MC-Glauber (dashed) and IP-Glasma (solid) model.}
     \label{fig:v3-fixedNpart-dAu}
   \end{center}
\end{figure}

We next present $v_2$ and $v_3$ as functions of transverse momentum $p_T$ for p+Pb collisions in Figs.\,\ref{fig:v2-fixedNpart} and \ref{fig:v3-fixedNpart} and for d+Au collisions in Figs.\,\ref{fig:v2-fixedNpart-dAu} and \ref{fig:v3-fixedNpart-dAu}. We see the same trend as observed for the integrated $v_n$.

Finally, we can qualitatively compare our results here to the results of the LHC and RHIC experiments on proton-nucleus and deuteron-nucleus collisions. The trend of $v_2$ as a function of centrality observed in p+Pb collisions appears to be different from that of the ALICE data~\cite{Abelev:2012ola} on proton-lead collisions at $\sqrt{s}=5.02$ TeV/nucleon. However, the error bars in the published data are too large to draw a definitive conclusion at present. The ATLAS collaboration has also presented~\cite{Aad:2013fja} a quantity called $v_2$({\rm {PC}), which is defined similarly to the ALICE $v_2$ and has the same trend as the ALICE results. However, the collaboration also presents results for 
$v_2\{4\}$, from four particle correlations, which appears to have the opposite trend with centrality relative to $v_2({\rm PC})$. The computations of~\cite{Bozek:2013uha} appear to be in agreement with this $v_2\{4\}$ quantity for the centralities compared. However, as can be seen in Fig.~\ref{fig:v2tot-pPb}, the IP-Glasma results are approximately a factor of two lower for the $N_{\rm part}$ that correspond to the same centrality selection. Note further that the IP-Glasma results are for $\eta/s=0.08$ and will be smaller for the $\eta/s =0.2$ that gives a good description of 
$v_2$ in A+A collisions at the LHC. We also note that the RHIC d+Au results on the ridge are reproduced by the MC Glauber 1 model. The differences between this model and the IP-Glasma model (for integrated $v_2$ values) in these collisions can be seen in Fig.~\ref{fig:v2tot-dAu}. 
More quantitative studies and additional data will clearly help clarify the role of hydrodynamics in the interpretation of the RHIC and LHC results on the ridge in deuteron-gold and proton-lead collisions respectively. 

\section{Validity of viscous hydrodynamics}

In the previous sections, we discussed the strong dependence of the initial geometry on initial state dynamics when one considers the especially small sized systems one has in proton/deuteron-nucleus collisions. We also discussed the hydrodynamic flow resulting from these initial spatial geometries. One may ask in addition under what conditions, if any, hydrodynamics is applicable to these especially small sized systems. Hydrodynamics is usually a good effective field theory in the late time, long wavelength limit of the theory. One way to quantify whether this holds in the systems of interest is to compare the relative magnitude of the viscous terms in the stress-energy tensor to the ideal fluid terms. Considering for simplicity only shear effects, and neglecting heat flow and bulk viscosity, one has 
\begin{equation}
T^{\mu\nu} = T_0^{\mu\nu} + \pi^{\mu\nu} \,,
\end{equation}
where the ideal term $T_0^{\mu\nu} = (e+P)g^{\mu\nu} - P g^{\mu\nu}$ and the viscous part of the stress-energy tensor satisfies the equation 
\begin{align}
\tau_\pi \Delta_\alpha^{~\mu}&\Delta_\beta^{~\nu} u^\lambda \partial_\lambda \pi^{\alpha\beta} + \pi^{\mu\nu} = -\frac{4}{3}\pi^{\mu\nu}(\partial_\lambda u^\lambda)\notag\\ 
 & + \eta \left[\left(\nabla^\mu u^\nu + \nabla^\nu u^\mu\right) - \frac{2}{3}\Delta^{\mu\nu}\nabla_\lambda u^\lambda\right]\, .
\end{align}
Here, $e$ is the energy density and $P$ the pressure, the metric $g^{\mu\nu} = {\rm diag}(1,-1,-1,-1)$, the transverse projector $\Delta^{\mu\nu} = g^{\mu\nu}-u^\mu u^\nu$ and $\nabla^\mu = \Delta^{\mu\nu}\partial_\nu$. The relaxation time is set to $\tau_{\pi} = 3\eta/(e + P)$.

We cannot initialize with the full $T^{\mu\nu}$ provided by the IP-Glasma model at the initial time, because of its highly non-equilibrium nature.
Therefore, we are left with a choice for $\pi^{\mu\nu}_0$ at the initial switching time $\tau=0.2\,{\rm fm}$ between flow in the Glasma and the later flow described by hydrodynamics. We study the case $\pi_0^{\mu\nu}=0$, as often employed in viscous hydrodynamic simulations, and the case where $\pi^{\mu\nu}_0$ takes on its Navier-Stokes value,
\begin{equation}
  \pi^{\mu\nu}_0 = \eta \left(\nabla^\mu u^\nu + \nabla^\nu u^\mu -\frac{2}{3}\Delta^{\mu\nu}\nabla_\lambda u^\lambda \right)\,.
\end{equation}
In this work we neglect gradients of the transverse velocities at the initial time and only keep the dominant piece from the longitudinal (boost-invariant) dynamics.

To avoid instabilities of the algorithm {\it outside the freeze-out surface}, where viscous corrections can become very large, we
introduce a regulator that restricts the viscous correction to be smaller than ten times the ideal part. In practice, we require
$
  \sqrt{\pi_{\mu\nu} \pi^{\mu\nu}} \leq 10 \sqrt{e^2+3P^2}
$
and implement a continuous regulation of $\pi^{\mu\nu}$ to satisfy this requirement:
$  \pi^{\mu\nu} \rightarrow \hat{\pi}^{\mu\nu} = \pi^{\mu\nu} \tanh(\rho)/\rho\,,$
where $\rho = \sqrt{\pi_{\mu\nu} \pi^{\mu\nu}}/(10 \sqrt{e^2+3P^2})$.
We have checked that the results presented for cells within the freeze-out surface are very weakly sensitive to this regulator
when the code is stable without regulation. The only effect of the regulator is to avoid instabilities that are triggered outside the freeze-out surface.

To quantify the validity of the viscous hydrodynamic approach for different collision systems, we can determine the fraction of cells within the freeze-out surface that have a viscous correction $\sqrt{\pi_{\mu\nu} \pi^{\mu\nu}}$ that are either larger than 25\% (Figs. \ref{fig:viscousFractionAvg025} and  \ref{fig:viscousFractionAvgNS025}) of the ideal fluid contribution $\sqrt{e^2+3P^2}$, or larger than a 50\% (Fig. \ref{fig:viscousFractionAvgNS05}) to the same. The ratio $\sqrt{\pi_{\mu\nu} \pi^{\mu\nu}}/\sqrt{e^2+3P^2}$ plays the role of an effective inverse Reynolds number~\cite{Baym:1985tna,Dumitru:2007qr}.

The results shown are for two different values of $\eta/s$, 0.08 and 0.2 and are for central ($b=0\,{\rm fm}$) Pb+Pb and central p+Pb collisions as a function of time $\tau$ for zero initial $\pi^{\mu\nu}$ 
in Fig.\,\ref{fig:viscousFractionAvg025}, and for an initial Navier-Stokes $\pi^{\mu\nu}$ in Figs.\,\ref{fig:viscousFractionAvgNS025} and \ref{fig:viscousFractionAvgNS05}.  The results are averages over 10 events each. We find very similar behavior in p+Pb and Pb+Pb collisions, with many cells having large corrections at times $\tau \lesssim 2\,{\rm fm}$, especially for the larger value of $\eta/s=0.2$.
In the latter case, which gives the best agreement with heavy ion collision data at the LHC~\cite{Gale:2013da}, nearly all cells have corrections of at least 25\% at early times, even for zero initial $\pi^{\mu\nu}$. In the case of Navier-Stokes initial conditions, all cells within the freeze-out surface start out with an at least 50\% viscous correction. For the smaller $\eta/s=0.08$, viscous corrections are small after $\tau=1\,{\rm fm}/c$ for both initial $\pi^{\mu\nu}$ choices. 

Although the results seem very similar for p+Pb and Pb+Pb collisions, it is important to note that the lifetime in Pb+Pb collisions is about 6 times longer than in p+Pb collisions. This means that viscous corrections are large for a significant fraction of the total lifetime of p+Pb collisions, while they are large only for a small fraction of the total lifetime of Pb+Pb collisions. In other words, depending on the value of $\eta/s$, viscous hydrodynamics in the first $1$-$2\,{\rm fm}/c$ is sensitive to not just the initial spatial geometry but also to details of how the viscous flow is initialized. While there is still a 
significant part of the space-time evolution where viscous hydrodynamics is a reliable framework for Pb+Pb collisions, the same cannot be said for p+Pb collisions. In this case, the system dilutes by 3-$4\,{\rm fm}/c$ -- thus, viscous hydrodynamics is unreliable for approximately half the lifetime of the 
system in p+Pb collisions, and therefore a source of potentially large systematic errors.  Similar conclusions and limits on the initial time $\tau_0$ when viscous hydrodynamics becomes applicable have been found in \cite{Dumitru:2007qr} for heavy-ion collisions.

\begin{figure}[htb]
   \begin{center}
     \includegraphics[width=8cm]{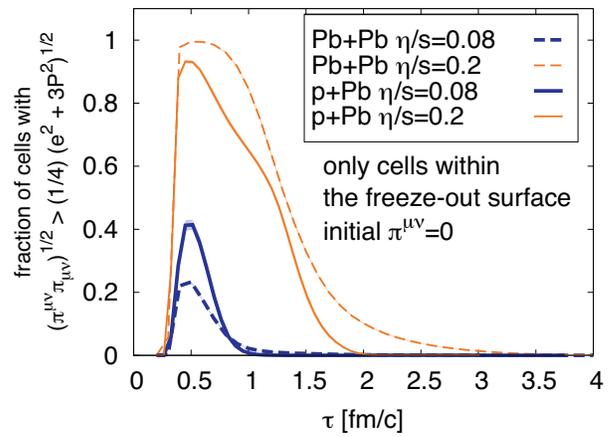}
     \caption{ (Color online) Fraction of all cells within the freeze-out surface that have a viscous correction of at least 25\% of the ideal $T^{\mu\nu}$. 
     Initialization with $\pi^{\mu\nu}_0=0$.}
     \label{fig:viscousFractionAvg025}
   \end{center}
\end{figure}

\begin{figure}[htb]
   \begin{center}
     \includegraphics[width=8cm]{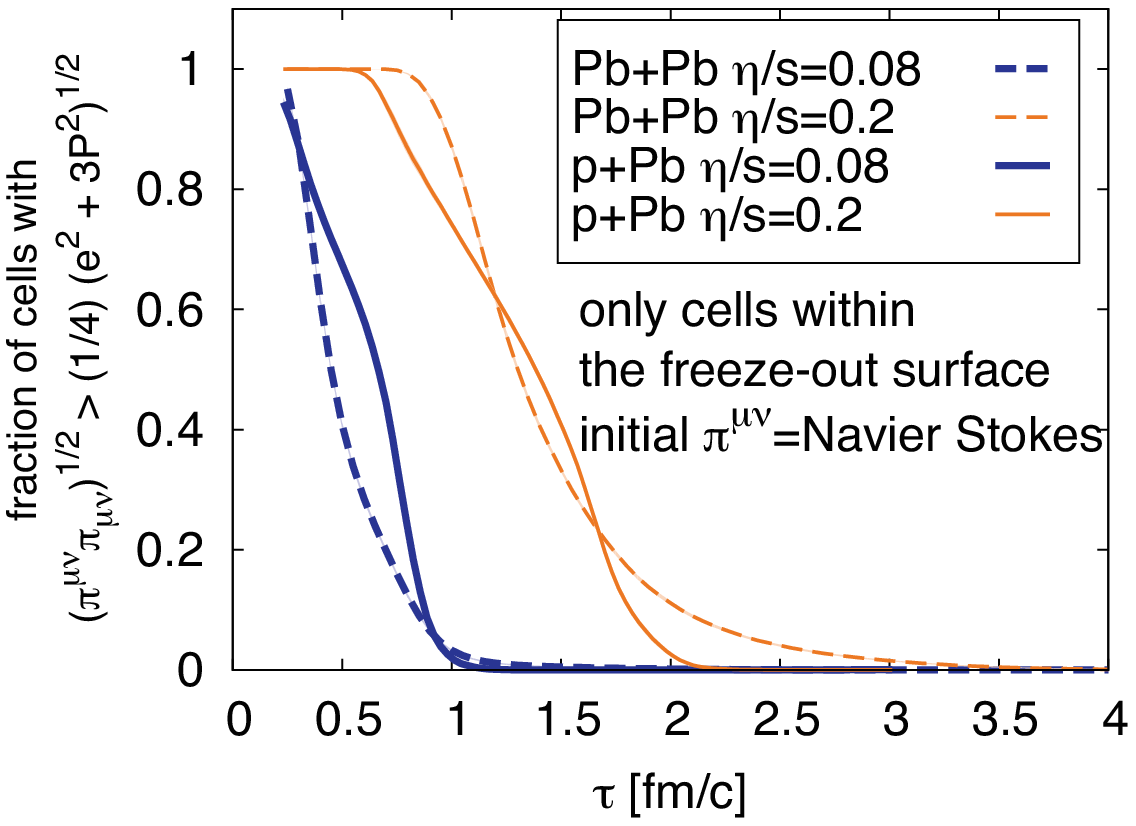}
     \caption{ (Color online) Fraction of all cells within the freeze-out surface that have a viscous correction of at least 25\% of the ideal $T^{\mu\nu}$. 
       Initialization with the Navier-Stokes value for $\pi^{\mu\nu}_0$.}
     \label{fig:viscousFractionAvgNS025}
   \end{center}
\end{figure}

\begin{figure}[htb]
   \begin{center}
     \includegraphics[width=8cm]{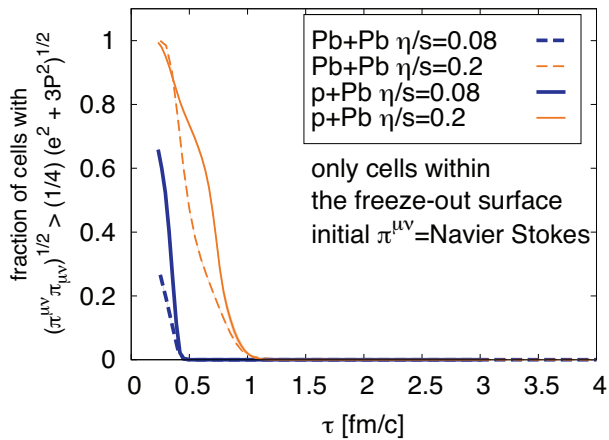}
     \caption{ (Color online) Fraction of all cells within the freeze-out surface that have a viscous correction of at least 50\% of the ideal $T^{\mu\nu}$. 
       Initialization with the Navier-Stokes value for $\pi^{\mu\nu}_0$.}
     \label{fig:viscousFractionAvgNS05}
   \end{center}
\end{figure}

\section{Summary and outlook} 

In this work, we have demonstrated that different dynamical assumptions can lead to qualitatively different results for spatial ellipticities and triangularities in smaller size systems. In contrast, the same range of dynamical assumptions give qualitatively comparable results in heavy ion collisions, with some differences showing up in quantitative studies. Results for the flow coefficients $v_2$ are significantly different between the MC-Glauber 1 model and the IP-Glasma model in proton-nucleus collisions at the LHC. The results studied are for $\eta/s=0.08$, and one expects that $v_2$ in both models will be significantly smaller for the value of $\eta/s=0.2$; this value best describes the LHC Pb+Pb data in the IP-Glasma model. The trend of $v_2$ with centrality seen in both models is different from the preliminary ALICE $v_2$ and ATLAS $v_2$({\rm PC}) data on p+Pb collisions. More precise data from all the LHC experiments can resolve this issue--the four particle correlations measured by ATLAS are an interesting step in this direction. With regard to deuteron-gold collisions at RHIC, the differences in $v_2$ between the MC-Glauber 1 and IP-Glasma model are smaller; again, both results would be smaller for the $\eta/s=0.12$ value that gives the best fit to RHIC Au+Au data in the IP-Glasma model. $v_3$, even for the small $\eta/s$ used here, is quite small at both RHIC and the LHC. 

In addition to details of the initial spatial geometry, flow in small size systems can have large viscous contributions for a significant fraction of the lifetime of the system. Within the framework of the IP-Glasma+\textsc{music} model, we showed for instance that for $\eta/s=0.2$, viscous corrections are as large as 50\% for times up to a fermi. In the Pb+Pb case, much of the flow is built up at later times when these viscous corrections are small, and second order viscous hydrodynamics can be considered a good effective description of the collective dynamics. In contrast, because the lifetime of the system in p+Pb collisions is much shorter than in Pb+Pb, the large viscous corrections over a significant fraction of the lifetime suggest second order viscous hydrodynamics is less reliable in such systems.

Hydrodynamics has provided us with a very successful framework to interpret the results of momentum anisotropies observed in heavy ion collisions. 
The recent experiments on proton/deuteron-nucleus collisions have brought to the fore the question of the applicability of hydrodynamics to these very small sized systems. The interest in further studies along the lines of this work is two-fold. Firstly, it would be important to quantify where hydrodynamics breaks down to establish a better understanding of the transport properties of the Quark-Gluon plasma. Here a careful study of the ratio 
$v_2/\varepsilon_2$ as a function of the transverse density, $N_{\rm ch}/S_\perp$, in various systems could be very useful. However, as we emphasized in this paper this problem is not trivial because in small systems various models lead to very different eccentricities.   

Secondly, these studies focus our attention on a better understanding of the initial non-equilibrium dynamics of strong color fields. For instance, if the azimuthal collimation of the nearside p+p, p+Pb  and d+Au ridge is not due to hydrodynamic flow, a subtle long range quantum interference between saturated gluon fields is likely the successful alternative explanation. On the other hand, if hydrodynamics is a viable explanation, quantitative comparisons to data can help narrow down the dynamics that generates the initial spatial geometries that subsequently experience hydrodynamic flow.

\emph{Acknowledgments}
We thank Adrian Dumitru, Kevin Dusling, Larry McLerran, Jamie Nagle, Zhi Qiu, Anne Sickles and Derek Teaney 
for interesting discussions. 
AB is supported through the RIKEN-BNL Research Center. BPS\ and RV\ are supported 
under DOE Contract No.DE-AC02-98CH10886.
BPS gratefully acknowledges a Goldhaber Distinguished Fellowship from Brookhaven Science Associates.

\vspace{-0.5cm}
\bibliography{spires}

\end{document}